# Harmonic Suppression Study on Twin Aperture CCT Type Superconducting Quadrupole for CEPC Interaction Region

PENG Quanling，XU Qingjin

*Abstract*—Iron free twin aperture superconducting quadrupole in the interaction region is a key technology to increase the luminosity of the high energy collider. CEPC that China will build in the next 10 years, has the center-of-mass energy of 240 GeV and small cross angle of 33 mrad in the interaction region. The final focusing quadrupole QD0, which is 2.2 m away from the interaction point, only have a 72 mm beam separation distance at the front end of the magnet. Because of the tight space, Canted Cosine Theta（CCT）coil type is the best selection, where the coil can be wound directly on the coil former in a fixed inclination angle. For QD0, the superconducting quadrupole coils for the two apertures are nearly contacted, high order harmonics will be produced by magnetic field crosstalk and should be canceled out at the design stage. This paper will present the design study for a pair of 400 mm long QD0 prototype, where we deliberately introduce some opposite harmonics to optimize the coil configuration so as to cancel out the unwanted harmonics. From the field calculation, we can also draw a conclusion that the different order harmonics have an independent property, which we can design a combined magnet by adding some high order harmonics, so that it can save space and reduce the magnet cost.

1. Introduction

One of the CEPC high-energy physics goals is to provide $e^+$-$e^-$ collisions at a beam energy of 120 GeV and attain a luminosity of $3\times10^{34}$ $cm^{-2}s^{-1}$ at each interaction  point (IP) when operated in the Higgs mode. It can also be able to run at the low energy as 80 GeV or 45.5 GeV for the W and Z experiments modes [1]. At the present design, all the CEPC storage ring, booster ring and the future SPPC storage ring which be housed in the same 100 km circumference tunnel. As shown in Fig. 1, for the CEPC double storage rings, there are four long straight sections, two for the beam interaction region (IR), two for superconducting RF cavities. In order to achieve the high luminosity, it requires multi-bunches, high beam current, small beam size at and small cross angle at IPs to realize the nearly head to head collision.

Fig. 2 shows the central part of the detector, where the horizontal crossing angle for the incoming electron and positron beams is only 33 mrads. The lattice design requires the focusing quadrupole QD0 with the length of 2.0 m to be 2.2 m away from the IP, which allows very tight space to separate the two quadrupoles. A reliable solution is to build the quadrupoles in a twin-aperture pattern, where the space between the two beam pipes can be fully used by the magnet coils.

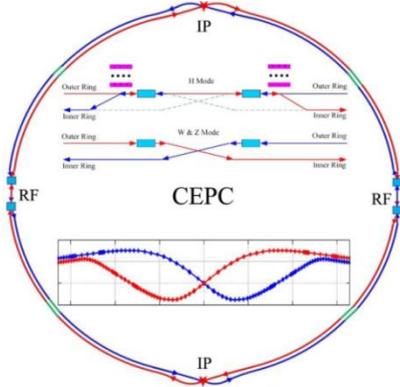

Fig. 1 Layout of the CEPC Collider.

Since the superconducting coils in each aperture are nearly contacted, it will give rise up to a strong crosstalk and should be reduced to an accepted level. As shown in Figure 2, QD0 and QF1 are all twin aperture quadrupoles, they are operated inside the detector with the central field of 3.0 T, similar as BEPCII [2][3]. A series of anti-solenoids and shield solenoids are needed before QD0 and surrounding QD0 and QF1 to reduce the coupling effect that coming from the detector solenoid field.

The work is supported by the Strategic Priority Research Program of the Chinese Academy of Sciences Grant No. XDB25000000, and the National Natural Science Foundation of China (Grant No. 11675193, 11575214, 11604335). Quanling Peng is with the Accelerator Research Center, Institute of High Energy Physics, Chinese Academy of Sciences, Beijing, 100049 China. (Tel: +86-10-88235904, E-mail: pengql@ihep.ac.cn).

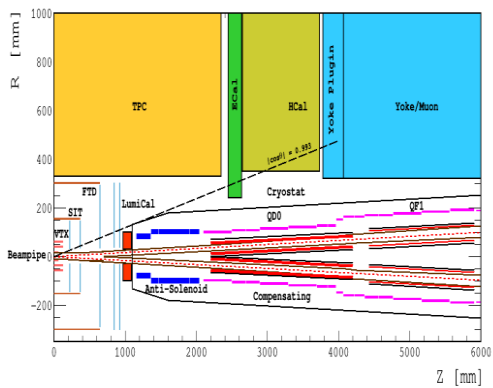

Fig. 2. Layout of the central part of the detector. Twin aperture quadrupole QD0 and QF1, three sets of compensation solenoids are contained in the same magnet cryostat.

I. Coil design for single aperture CCT superconducting quadrupole

In recent years, with the improvement of CNC lathe and 3D printing technology, CCT type coil that proposed in 1970s [4] are mentioned again many laboratories and find applications in particle accelerator [4][5][6][7]. The advantages of using CCT coils are including the followings:

1). CCT coils can reach an excellent field quality. A pure dipole or quadrupole field can be produced according the standard spatial function without need special adjustment for the coil end. 2). Simplicity in engineering design, because of low current operation and low electromagnetic force, so no coil collars are needed, which is simpler and cheaper than conventional using Rutherford cable. The CCT coils are embedded into the slots that pre-machined on the coil former. By epoxy impregnation, the main body of the superconducting coil can withstand the electromagnetic forces. 3). An important application, the CCT coils also can be designed with the combination of the several function magnets, or can be used for harmonic correction.

The disadvantage for using CCT coil lies in: one coil layer can produce a chosen multipole field plus an extra solenoid field, which needs another coil layer to cancel out the solenoid field and add up the designed fields. Compare with the conventional cosine theta coil design, it needs more superconducting conductor.

Because of the separation distance at front end of QD0 magnet is only 72 mm, the first 400 mm long coils can take only one double layer. Going from 2.6 m to 4.2 m, when the beam separation space increasing, CCT coil in each aperture can take two double layers, so the total integral field can meet the physical requirements.

If a single strand wire goes along the spatial function as Eq. 1 [7][8], a pure quadrupole field plus a solenoid field can be produced. Here R is the mean radius of the coil layer, $\theta$ is the space angle for each turn, $\omega$ is the coil twist pitch, $\alpha$ is coil winding skew angle with respect to the coil axis. To produce a pure quadrupole field, we should put another layer about lager R, with the negative symbol for the first term in $z$ for Eq. 1. With that, the solenoid field from two layers can cancel out, but the quadrupole field added up.

$$x = R\cos\theta$$
$$y = R\sin\theta \quad\quad\quad (1)$$
$$z = \left(\frac{R\sin(2\theta)}{2\tan\alpha} + \frac{\omega\theta}{2\pi}\right)$$

In 2D space, magnet field inside the magnet aperture can be written as Eq. (2)[3], Here $a_n$ and $b_n$ are respectively the shew and normal components, they are usually normalized with fundamental field, and expressed in units (1 unit=$1\times 10^{-4}$). For a quadrupole field n=2, we expected the scale of the high order harmonics (from n=3 to n=10) less than 10 unit, here the dominant quadrupole field $B_2$ is set to 1, with $B_2 = \sqrt{a_2^2 + b_2^2}$. $R_0$ is the reference radius, normally it is the 2/3 of the beam pipe. The fundamental and high order harmonics in each cross section can be calculated by FFT. For a 3D field, harmonic for each order n can be calculated by summing up the corresponding value in every cross section along the beam line, and then normalized to the fundamental quadrupole $B_2$.

$$B_x + iB_y = 10^{-4} B_2 \sum_{n=1}^{\infty}(a_n + ib_n)(\frac{x+iy}{R_0})^{n-1}. \qquad (2)$$

For easy coil winding, totally 8 NbTi wires with each diameter of 0.82 mm are bounded together in a 2×4 shape, is put into the pre-machined slot in the coil former. Table 1 shows the parameters for 400 mm CCT Quadrupole prototype.

OPERA-3d [9] was used for the magnetic field calculation, the cable is taken as a $2\times 4$ mm$^2$ block for the coil model, high order field harmonics relative to the quadrupole components are taken out by FFT at the reference radius $R_0$=13 mm. When the operation current is 600 A for a single strand, the calculated field gradient is 67 T/m, which is less than the required field gradient of 130 T/m. As it is only one double layer for the 0.4m long prototype, add up the following two double layers of 1.6 m long, the total integral field strength can meet the physical requirements. It is the same idea as CERN reported in 2015 for an 18 T Hybrid CCT dipole [10].

Table 1 Parameters for 400 mm CCT Quadrupole prototype

| Symbol | unit | value |
| --- | --- | --- |
| Strand diameter | mm | 0.825 |
| Ic(@5T, 4.2K) | A | 620 |
| Numbers of turns per slot |  | 2×4 |
| Slot size on the coil former | mm | 2.1×4.2 |
| Numbers of layers |  | 2 |
| Coil frame thickness | mm | 5.2 |
| Canted coil angle | degrees | 30 |
| Axial pitch length ω | mm | 5.2 |
| Design gradient G | T/m | 67 |
| Inner diameter of the inner layer | mm | 46 |
| Inner diameter of the outer layer | mm | 58 |
| Beam pipe inner diameter | mm | 40 |
| Peak field on the coil | T | 2.3 |

Fig. 3 shows coil configuration for the 400 mm CCT prototype. From upper to lower, they are respectively the inner coil, the outer coil and two upper coils combination that produce a set of CCT coils.

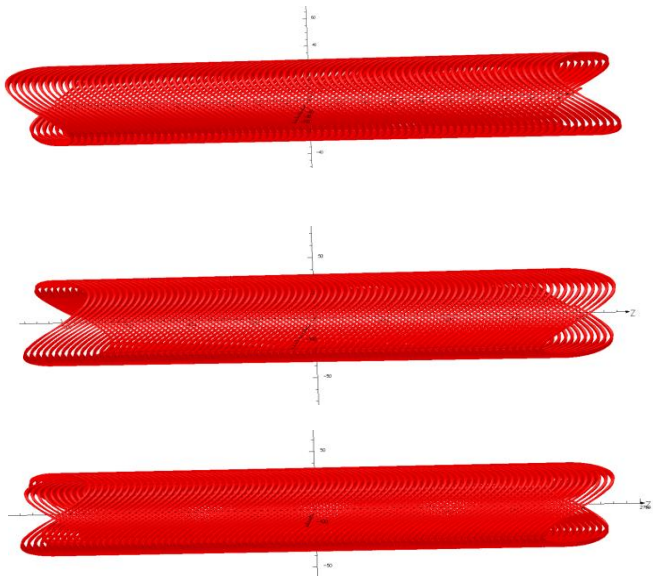

Fig. 3. 400mm CCT quadrupole coil prototype. Upper: the inner coil; middle: the outer coil, lower: two coil layers put together to form a set of the CCT coil.

Fig. 4 shows the harmonic distribution along the beam line, they are normalized to the quadrupole field at reference radius of R=13 mm. It can be seen, all the high order harmonics in the central part are nearly to zero, n=3 and other higher harmonics are all less than 1 unit, even at the coil end. As can be seen, the integral field for each high order harmonic is nearly to zero, since they are canceled out at the coil ends. The data for the harmonics higher than n=6, which are very small, are shown in the figures.

II. Harmonic generated by field crosstalk

If put the two sets of CCT coils in closer, magnetic fields in the two apertures will disturb from each other and bring high order harmonics. Fig. 5 shows the two sets of CCT coil that situated as the similar position as that of to QD0. Fig. 6 and Fig. 7 respectively show the harmonic distribution in each aperture 1 (AP1) and aperture 2 (AP2). The calculated harmonics are shown in the second and the fifth column in Table 2. It can be seen, a sextupole (n=3) and octupole (n=4) components increase higher than the physical requirements, the values are nearly 100 units, which need to cancel out.

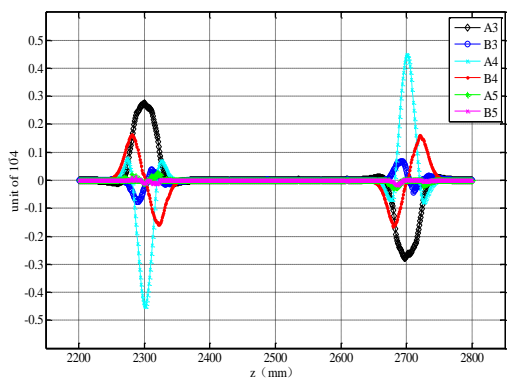

Fig. 4. Harmonic distribution for the 400mm pure quadrupole coil(@R=13mm) as shown in Fig. 3. The $z$ axis is along the beam line, the $y$ scale is in unit.

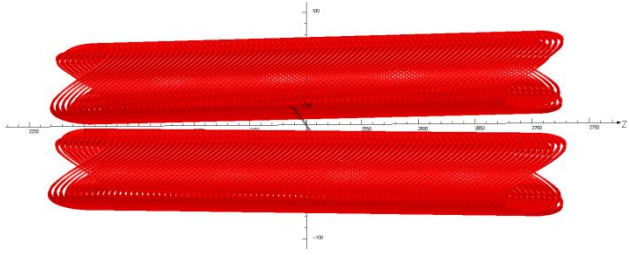

Fig. 5. Two 400 mm CCT coil models are put in the similar position as that of QD0.

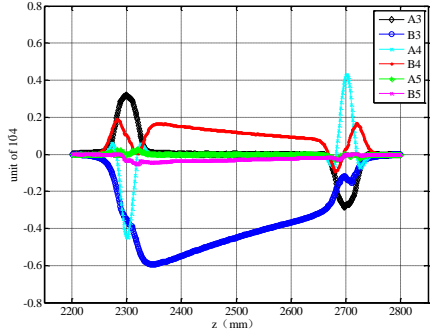

Fig. 6. Harmonic distribution along the beam axis in AP1.

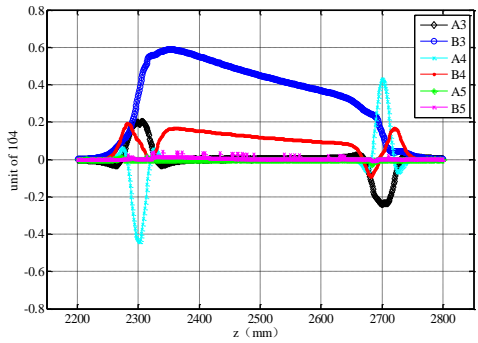

Fig.7. Harmonic distribution along the beam axis in AP2.

III. Introduced opposite harmonics in the CCT coils

In order to cancel the harmonics generated by field crosstalk, several extra $z$ terms will be deliberately added in Eq. 1, they are shown in Eq. 3 [3][8]. Noted the $C_m$ and $D_m$ are the cancelation factors for harmonic order $m$, they are used to cancel the normal and skew field harmonics respectively.

$$z = \sum_{m_b}(C_m \frac{R\sin(m_b\theta)}{m_b \tan\alpha}) + \sum_{m_a}(D_m \frac{R\cos(m_a\theta)}{m_a \tan\alpha}) \qquad (3)$$

As the known orthogonality property, when m≠n, $\int_0^{2\pi} \sin(m\theta)\sin(n\theta)\,d\theta = 0$, or $\int_0^{2\pi} \sin(m\theta)\cos(n\theta)\,d\theta = 0$, the new added $m$ order harmonics do not affect the scale of the fundamental quadrupole field and other order harmonics. The modified coil configuration can be rebuilt according to Eq. 1 and Eq. 3. For AP1 CCT coils, we introduce the opposite $b_3$, $b_4$ and $b_5$ values with the same absolute values in the second column in Table 2, and do the calculation again by Opera-3d. The calculated results are shown in the third column. It can be seen the new generated harmonics are almost as we added, noted that other harmonics which are less than 1 unit are not shown for clear comparison. Same method is taken for AP2, the calculated results are shown in the sixth column in Table 2.

Table. 2. High order harmonics before and after field correction (unit: $10^{-4}$)

|  | AP1 | | | AP2 | | |
| --- | --- | --- | --- | --- | --- | --- |
|  | Cross talk | Added harmonics | After correction | Crosstalk | Added harmonics | After correction |
| b3 | -185 | 187 | -1.70 | 186 | -187 | 0.32 |
| a3 | 2.14 | ---- | 2.56 | 2.31 | --- | -2.21 |
| b4 | 50 | -52 | -2.49 | 50 | -46 | 3.59 |
| a4 | -0.86 | ---- | -0.96 | -0.75 | --- | -0.68 |
| b5 | -12 | 10.8 | -1.56 | 2.61 | --- | 1.90 |
| a5 | 0.15 | ---- | 0.26 | -0.064 | --- | -0.086 |
| b6 | 2.65 | ---- | 2.43 | 2.65 | --- | 2.33 |
| a6 | -0.05 | ---- | -0.034 | 0.056 | --- | 0.067 |

Fig. 8 and Fig. 9 show respectively the harmonic distribution in each aperture after added the opposite sextupole and octupole components. Noted that they are calculated in a single aperture state, no field crosstalk.

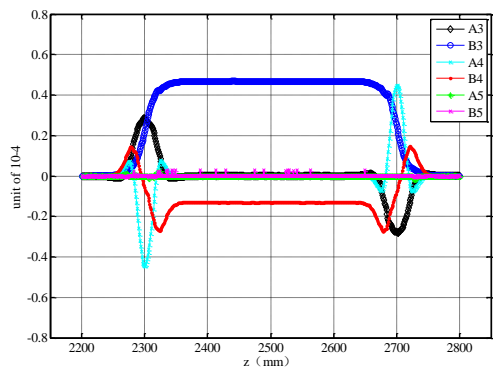

Fig. 8. Harmonic distribution in AP1 when added three opposite harmonics.

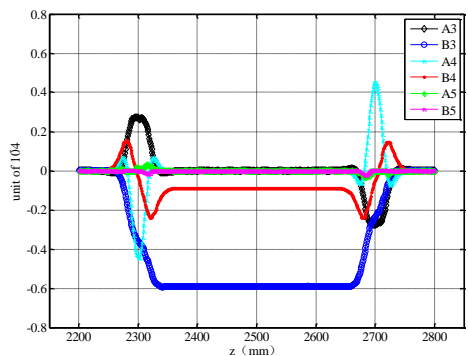

Fig. 9. Harmonic distribution in AP2 when added two opposite harmonics.

IV. Combined field calculation after introduced harmonics correction

Combined field was calculated after putting the modified coil configurations of AP1 and AP2 in the same position as shown in Fig. 5, the purpose is to check the harmonic suppression effects after added the opposite harmonics in each aperture. Fig. 10 and Fig. 11 show the suppressed harmonic distribution in AP1 and AP2 respectively. Compare with Fig. 6 and Fig. 7, the shape and scale for sextupole and octupole are greatly reduced. Their corresponding values are shown in the fourth column for AP1 and the seventh column for AP2 in Table 2; where the harmonics in each aperture are reduced to an accepted level. As QF0 is an iron free quadrupole, if we do not care about the persistent current effect, the field harmonics can keep the same value along with the current. For a single aperture coil, AP1 or AP2, whatever it is in room temperature or in 4.2 K state, the added harmonics can be measured by a long rotating coil, which can check the scale of the harmonics we added. On the other hand, when put the coils for AP1 and AP2 in the similar position as that of QD0, their harmonic cancelation effect can be checked out both at room temperature or in cryogenic state.

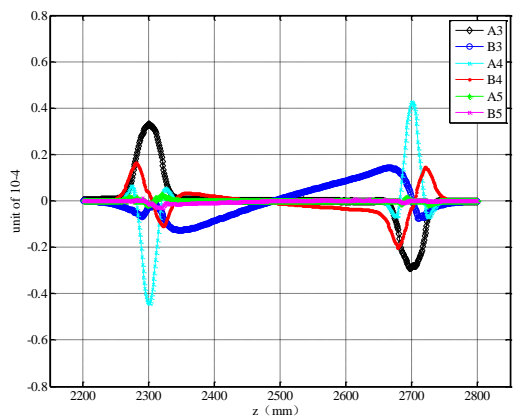

Fig.10. Harmonic distribution in AP1 after harmonic correction.

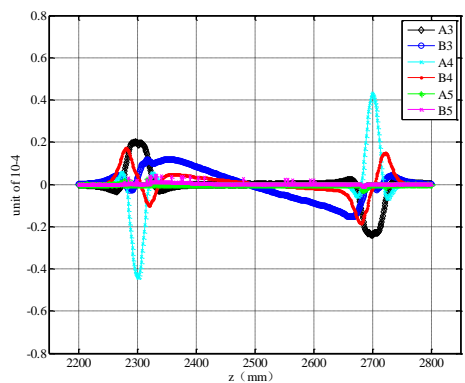

Fig. 11. Harmonic distribution in AP2 after harmonic correction.

V. Conclusion and prospect

Two sets of 400 mm CCT quadrupole coils are designed to verify the possibility to fabricate a 2.2 m long twin aperture quadrupole in CEPC interaction region. Since the CCT coils in each aperture are very closer, they will bring strong crosstalk effect. By adding opposite harmonics in each modified coil configuration, the unwanted harmonics can be eventually canceled out in each aperture. In the future, two sets of 400 mm CCT coils will be fabricated and put them in the same relative position as that of QF0, then perform the field measurement both at room temperature and at 4.2 K to test the reliability of coil design and harmonic correction effect. Field orthogonality property will lead us to design a multifunction CCT magnet, which can save space and reduce the total cost of the magnet production.